\documentclass[12pt]{article}

\usepackage{scicite}
\usepackage{graphicx}
\usepackage{amsfonts}
\usepackage{sidecap}

\usepackage{times}

\newenvironment{sciabstract}{%
\begin{quote} \bf}
{\end{quote}}

\newcounter{lastnote}

\title{The Hidden Mass and Large Spatial Extent of a Poststarburst
  Galaxy Outflow}

\author
{Todd M. Tripp,$^{1\ast}$ Joseph D. Meiring,$^{1}$ J. Xavier
  Prochaska,$^{2}$\\
Christopher N. A. Willmer,$^{\ 3}$ J. Christopher Howk,$^{4}$ Jessica K. Werk,$^{2}$\\
 Edward B. Jenkins,$^{5}$ David V. Bowen,$^{5}$ Nicolas Lehner,$^{4}$\\
Kenneth R. Sembach,$^{6}$ Christopher Thom,$^{6}$ and Jason Tumlinson$^{6}$\\
\\
\footnotesize{$^{1}$Department of Astronomy, University of Massachusetts, Amherst, MA 01003, USA}\\
\footnotesize{$^{2}$UCO/Lick Observatory, University of California, Santa Cruz, CA 95064, USA}\\
\footnotesize{$^{3}$ Steward Observatory, University of Arizona,
  Tucson, AZ 85721, USA}\\
\footnotesize{$^{4}$Department of Physics, University of Notre Dame, Notre Dame, IN 46556, USA}\\
\footnotesize{$^{5}$Princeton University Observatory, Princeton, NJ 08544, USA}\\
\footnotesize{$^{6}$Space Telescope Science Institute, Baltimore, MD 21218, USA}\\
\\
\footnotesize{$^\ast$To whom correspondence should be addressed; E-mail:  tripp@astro.umass.edu.}
}

\date{}

\begin{document} 




\maketitle


\begin{sciabstract}
Outflowing winds of multiphase plasma have been proposed to regulate
the buildup of galaxies, but key aspects of these outflows have not
been probed with observations.  Using ultraviolet absorption
spectroscopy, we show that ``warm-hot'' plasma at 10$^{5.5}$ K
contains 10-150 times more mass than the cold gas in a poststarburst
galaxy wind. This wind extends to distances $>$68 kiloparsecs, and at
least some portion of it will escape. Moreover, the kinematical
correlation of the cold and warm-hot phases indicates that the
warm-hot plasma is related to the interaction of the cold matter with
a hotter (unseen) phase at $\gg 10^6$ K.  Such multiphase winds can
remove substantial masses and alter the evolution of poststarburst
galaxies.
\end{sciabstract}




Galaxies do not evolve in isolation.  They interact with other
galaxies and, more subtly, with the gas in their immediate
environments.  Mergers of comparable-mass, gas-rich galaxies trigger
star-formation bursts by driving matter into galaxy centers, but
theory predicts that such starbursts are short-lived: the central gas
is rapidly driven away by escaping galactic winds powered by massive
stars and supernova explosions or by a central supermassive black hole
\cite{hopkins06}. Such feedback mechanisms could transform gas-rich
spiral galaxies into poststarburst galaxies \cite{zabludoff96} and
eventually into elliptical-type galaxies with little or no star
formation \cite{synder11}. Mergers are not required to propel galaxy
evolution, however. Even relatively secluded galaxies accrete matter
from the intergalactic medium (IGM), form stars, and drive matter
outflows into their halos or out of the galaxies entirely
\cite{keres09,opp10}. In either case, the competing processes of gas
inflows and outflows are expected to regulate galaxy evolution.

Outflows are evident in some nearby objects
\cite{veilleux05,heckman00,rupke05,martin06} and are ubiquitous in
some types of galaxies
\cite{pettini01,tremonti07,steidel10,rubin10a,hamann99,grimes09};
their speeds can exceed the escape velocity. Nevertheless, their
broader impact on galaxy evolution is poorly understood. First, their
full spatial extent is unknown. Previous studies
\cite{veilleux05,martin06,rubin10b,moe09,dunn10,edmonds11,hamann11,kriss11,capellupo11}
have revealed flows with spatial extents ranging from a few parsecs up
to $\approx 20$ kiloparsecs (kpc). However, because of their low
densities, outer regions of outflows may not have been detected with
previously used techniques, and thus the flows could be much
larger. Second, the total column density and mass of the outflows is
poorly constrained. Previous outflow observations are often limited to
low-resolution spectra of only one or two ions (e.g., Na~{\sc i} or
Mg~{\sc ii}) or rely on composite spectra that cannot yield precise
column densities. Without any constraints on hydrogen (the vast bulk
of the mass) or other elements and ions, these studies are forced to
make highly uncertain assumptions to correct for ionization, elemental
abundances, and depletion of species by dust.  Finally, galactic winds
contain multiple phases with a broad range of physical conditions
\cite{veilleux05}, and wind gas in the key temperature range between
10$^{5} - 10^{6}$ K (where radiative cooling is maximized) is too cool
to be observed in X-rays; detection of this so-called ``warm-hot''
phase requires observations in the ultraviolet (UV).

To study the more extended gas around galaxies, including regions
affected by outflows, we used the Cosmic Origins Spectrograph (COS) on
the Hubble Space Telescope (HST) to obtain high-resolution spectra of
the quasistellar object (QSO) PG1206+459 ($z_{\rm QSO} = 1.1625$).  By
exploiting absorption lines imprinted on the QSO spectrum by
foreground gaseous material, we can detect the low-density outer
gaseous envelopes of galaxies, regions inaccessible to other
techniques. We focus on far-ultraviolet (FUV) absorption lines at rest
wavelengths $\lambda _{\rm rest} < 912$ \AA.  This FUV wavelength
range is rich in diagnostic transitions \cite{verner94}, including the
Ne~{\sc viii} 770.409, 780.324\AA\ doublet, a robust probe of warm-hot
gas, as well as banks of adjacent ionization stages.  The sight line
to PG1206+459 pierces an absorption system, at redshift $z_{\rm abs} =
0.927$, that provides insights about galactic outflows. This
absorber has been studied before \cite{ding03}, but previous
observations did not cover Ne~{\sc viii} and could not provide
accurate constraints on {\sc H~i} in the individual absorption
components.

\begin{figure}
\includegraphics[width=19.0cm, angle=0]{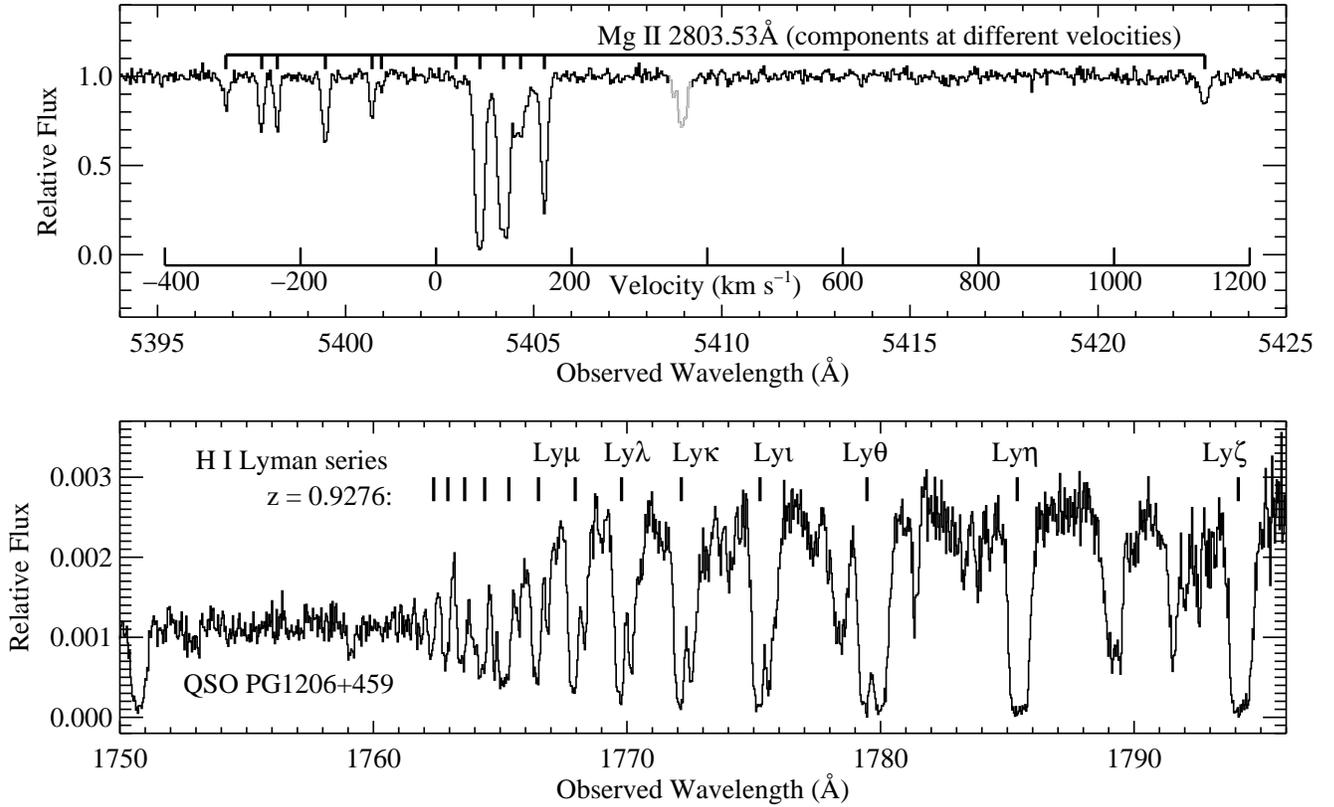} 
\caption{(Top) Small portion of the Keck HIRES spectrum of
  PG1206+459 \cite{ding03}. Tick marks at top indicate components detected at
  various velocities in the Mg~\textsc{ii} 2803.53 \AA\ transition. A
  velocity scale in the rest frame of the affiliated galaxy 177\_9 is
  inset at bottom.  Gray indicates a feature not due to Mg~\textsc{ii}
  2803.53 \AA . (Bottom) Small portion of the ultraviolet spectrum
  of PG1206+459 recorded with the Cosmic Origins Spectrograph (COS) on
  HST that shows {\sc H i} Lyman series absorption lines (marked with
  ticks and labels) at the redshift of the Mg~\textsc{ii} complex in
  the upper panel, including H~\textsc{i} Ly$\zeta$ through Ly$\sigma$
  (highest lines are marked but not labeled). \label{fig_llplot}}
\end{figure}

\begin{figure}
\includegraphics[width=19.0cm, angle=0]{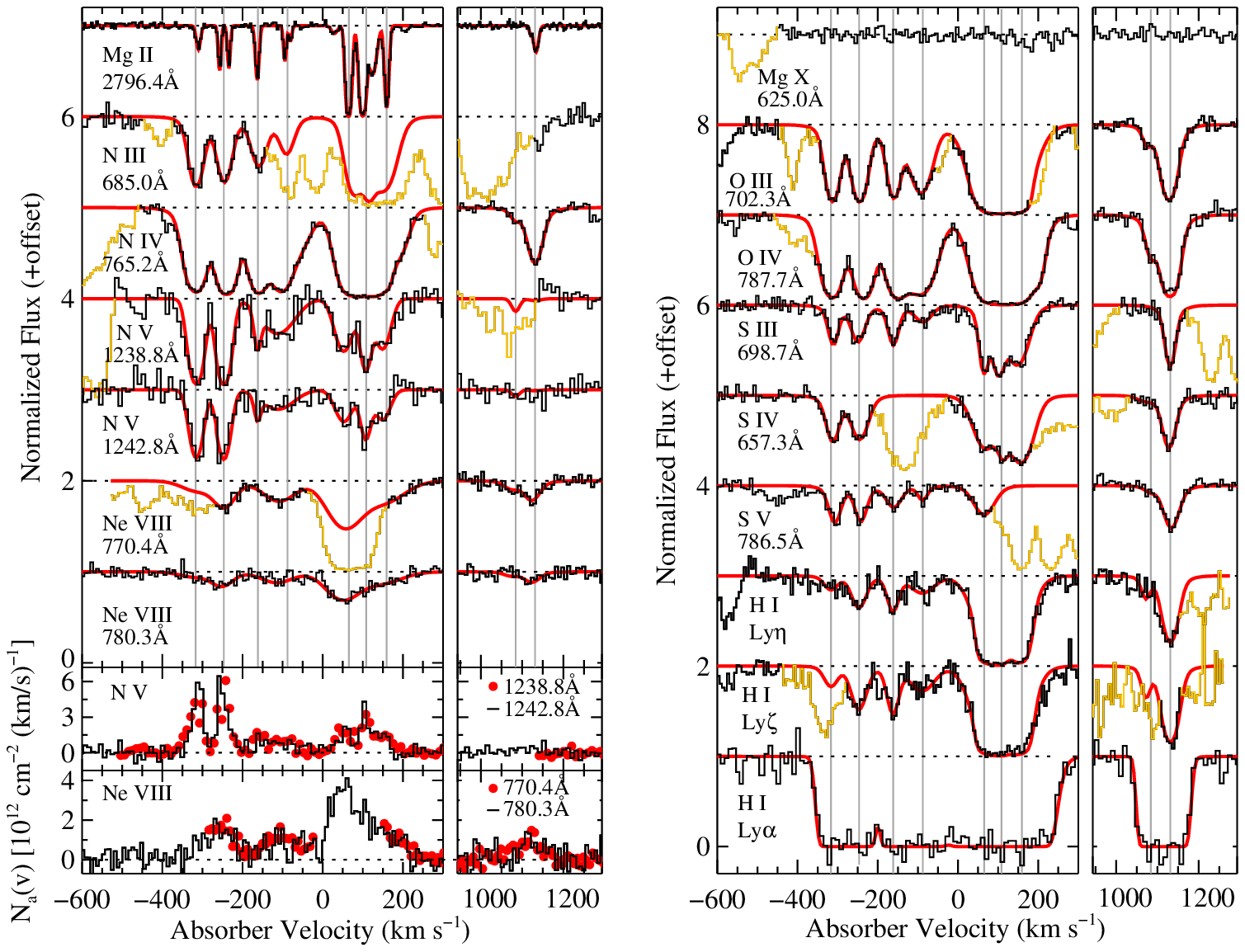}
\caption{Continuum-normalized absorption profiles (black
  lines) of various species detected in the Lyman limit/Mg~\textsc{ii}
  absorber shown in Figure~\ref{fig_llplot}, plotted in velocity with
  respect to the galaxy 177\_9 redshift (i.e., $v = 0$ km s$^{-1}$ at
  $z$ = 0.927).  Labels below each absorption profile indicate the
  species and rest wavelength. We fitted 9 components to the COS and
  STIS data \cite{ding03}. Component centroids are indicated by gray
  lines, and the Voigt-profile fits are overplotted with red lines
  \cite{refsom}. Yellow lines indicate contaminating features from
  other redshifts/transitions. The two panels at lower left compare
  apparent column density profiles \cite{savage91} of the {\sc N~v}
  and Ne~{\sc viii} doublets.\label{spec_stack}}
\end{figure}

\begin{figure}
\includegraphics[width=15.0cm, angle=0]{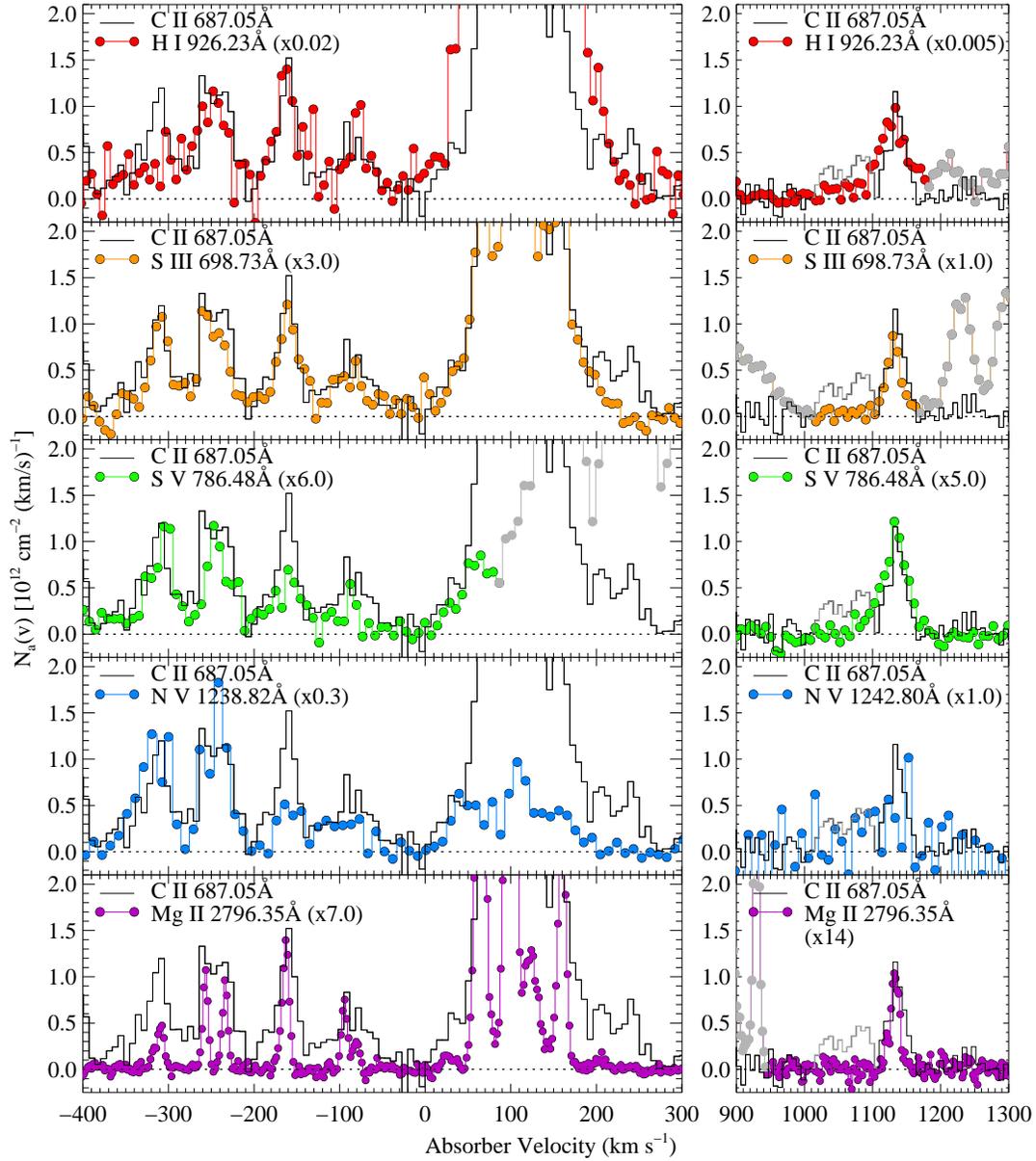}
\caption{Comparison of apparent column density profiles
  \cite{savage91} of the Lyman limit absorber affiliated with galaxy
  177\_9. In each panel, the C~\textsc{ii} 687.05 \AA\ profile (black
  histogram) is compared to another species (colored circles) as
  labeled at upper left; the comparison species profile is also scaled
  by the factor in parentheses following the species label. Gray lines
  indicate regions contaminated by unrelated absorption. As in
  Figure~\ref{spec_stack}, $v = 0$ km s$^{-1}$ at $z$ =
  0.927.\label{navcompare}}
\end{figure}

This absorber is illustrated in
Figures~\ref{fig_llplot}-\ref{navcompare} including the COS data
\cite{refsom}.  The absorber is a ``partial'' Lyman limit (LL) system
(i.e., the higher Lyman series lines are not saturated), which enables
accurate $N$({\sc H~i}) measurement (Fig.~\ref{fig_llplot}). A wide
variety of metals and {\sc H~i} lines are detected in at least 9
components \cite{refsom} spanning a large velocity range from $-317$
to $+1131$ km s$^{-1}$ (Figs.~\ref{fig_llplot}-\ref{spec_stack}). The
Ne~{\sc viii} doublet is unambiguously detected
(Fig.~\ref{spec_stack}) with a total Ne~{\sc viii} column density of
$N$(Ne~{\sc viii}) = $10^{14.9}$ cm$^{-2}$ \cite{refsom}, which is
$\approx 10\times$ higher than any previous $N$(Ne~{\sc viii})
detections in intervening absorbers \cite{savage05,narayanan11}. The
component at +1131 km s$^{-1}$ exceeds $v_{\rm escape}$ of any
individual galaxy, and the other components have very similar
properties to the +1131 km s$^{-1}$ component \cite{refsom},
suggesting a common origin. Whether the other components have $v >
v_{\rm escape}$ depends on the (unknown) potential well, but allowing
for projection effects and noting that the gas is already far from the
affiliated galaxy (see below), several of the other components could
also be escaping.  Combined with detection of Ne~{\sc viii}, the
detections of banks of adjacent ions ({\sc N~ii}/{\sc N~iii}/{\sc
  N~iv}/{\sc N~v};{\sc O~iii}/{\sc O~iv};{\sc S~iii}/{\sc S~iv}/{\sc
  S~v}) place tight constraints on physical conditions of the gas.
Notably, the velocity centroids and profile shapes of lower and higher
ionization stages are quite similar (Figure~\ref{navcompare}).

\begin{figure}
\includegraphics[width=15.0cm, angle=0]{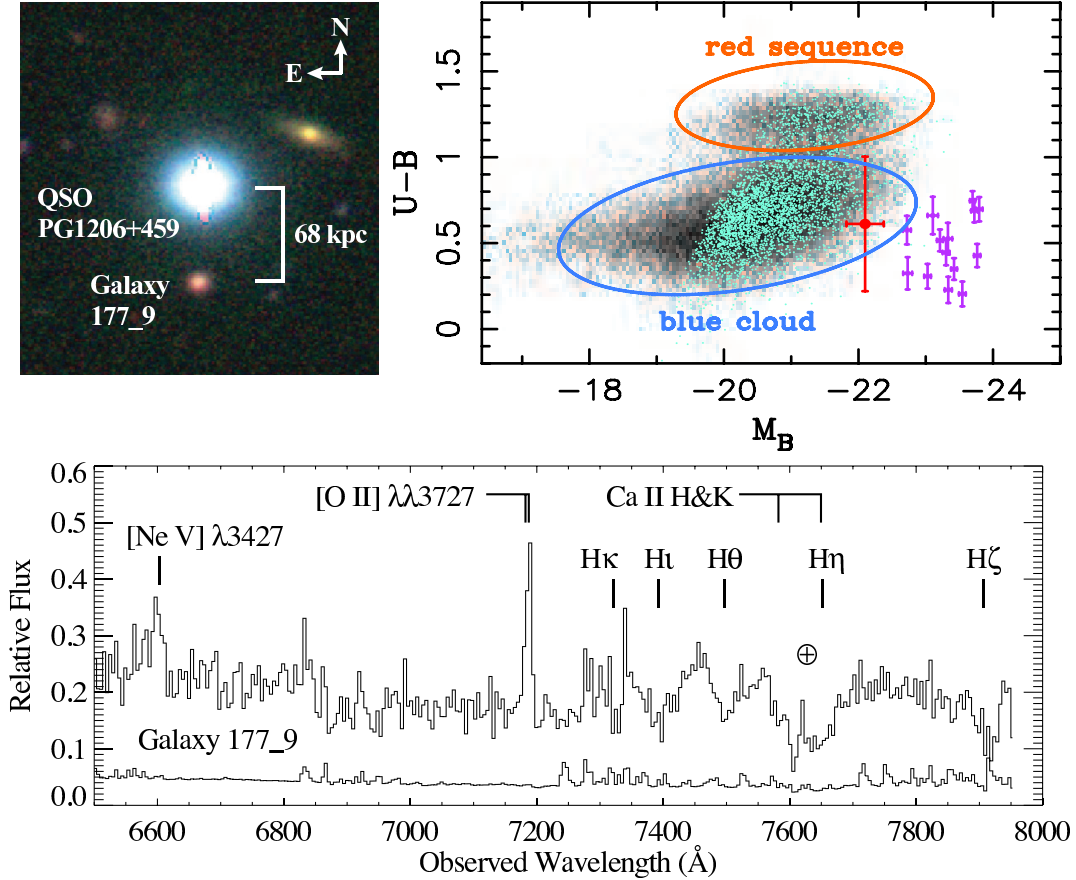}
\caption{Montage of observations of the galaxy at $z_{\rm
    gal} = 0.927$ that drives a large-scale outflow of metal-enriched
  plasma. The galaxy, and the background QSO which reveals the outflow
  via absorption spectroscopy, is shown in the upper-left panel in a
  multicolor image obtained with the Large Binocular Telescope.  This
  galaxy, which we refer to as 177\_9, is the red object 8.63
  arcseconds south of the bright QSO PG1206+459 ($z_{\rm QSO} =
  1.1625$) at a position angle of 177$^{\circ}$ (N through E) from the
  QSO. At the galaxy redshift, the angular separation from the QSO
  sight line corresponds to an impact parameter of 68 kpc. The large
  red circle in the upper-right panel indicates the rest-frame U-B
  color and absolute B magnitude of 177\_9 compared to all galaxies
  from the DEEP2 survey (grayscale, {\it 30}) and DEEP2 galaxies within
  $\pm$0.05 of $z$(177\_9) (cyan points).  The small purple circles
  show poststarburst galaxies from \cite{tremonti07}. The lower panel
  shows an MMT optical spectrum of 177\_9 (upper trace) with its
  $1\sigma$ uncertainty (lower trace). The strong feature at $\approx
  7600$ \AA\ is partially due to telluric
  absorption.\label{galaxy_montage}}
\end{figure}

This strong Ne~{\sc viii}/LL absorber is affiliated with a galaxy near
the QSO sight line \cite{refsom,ding03}. This galaxy, which we refer
to as 177\_9, is the type of galaxy expected to drive a galactic
superwind (Fig.~\ref{galaxy_montage}). Like poststarburst
\cite{tremonti07} and ultraluminous infrared galaxies \cite{chen10},
galaxy 177\_9 is very luminous and blue \cite{refcolor} -- based on M*
from DEEP2 \cite{willmer06}, the galaxy luminosity $L = 1.8 L$*. The
MMT spectrum in Fig.~\ref{galaxy_montage} is also similar to those of
the poststarburst galaxies in \cite{tremonti07}, with higher Balmer
series absorption lines, [O~\textsc{ii}] emission, and [Ne~\textsc{v}]
emission indicative of an AGN \cite{refsom}. Most importantly, the
galaxy has a large impact parameter from the QSO sight line, $\rho$ =
68 kpc \cite{cosmology}, which implies that the gaseous envelope of
177\_9 has a large spatial extent.

The component-to-component similarity of the absorption lines
(Fig.~\ref{navcompare}) suggests a related origin.  To further
investigate the nature of this absorber, we have used photoionization
models \cite{ferland98} to derive ionization corrections and elemental
abundances \cite{refsom}. These models indicate that the individual
components have high abundances ranging from $\approx 0.5\times$ to
$3\times$ those in the Sun (Table S2). Such high abundances (or
metallicities) favor an origin in outflowing ejecta enriched by
nucleosynthesis products from stars; at the large impact parameter of
177\_9, corotating outer-disk/halo gas or tidal debris from a low-mass
satellite galaxy would be expected to have much lower
metallicity. Tidal debris from a massive galaxy could have high
metallicity, but we are currently aware of only one luminous galaxy
near the sight line at the absorber redshift \cite{neargalnote};
another luminous galaxy interacting with 177\_9 is not evident. The
absorber could also be intragroup gas, but somehow it must have been
metal enriched, so some type of galactic outflow is implicated in any
case.

The photoionization models also constrain the total hydrogen column
(i.e., {\sc H~i} + {\sc H~ii}), and combined with $r \geq 68$ kpc,
this allows mass estimates.  Using fiducial thin-shell models
\cite{refsom}, we find that the mass of cool, photoionized gas in
individual components ranges from $0.6\times 10^{8} - 14\times 10^{8}$
solar masses.  However, photoionization fails (sometimes by orders of
magnitude) to produce enough S~\textsc{v}, N~\textsc{v}, and
Ne~~\textsc{viii}; these species must arise in hot gas at $T > 10^{5}$
K.  Using equilibrium and non-equilibrium collision ionization models
\cite{refsom}, we find that the warm-hot gas contains much more mass
than the cold gas, with individual components harboring $10\times
10^{8} - 400\times 10^{8} M_{\odot}$ in hot material.  These are rough
estimates with many uncertainties.  For example, if the absorption
arises in thin filaments analogous to those seen in starburst galaxies
\cite{veilleux05} or AGN bubbles \cite{fabian08}, the cold-gas mass
could reduce to $\approx 10^{6} M_{\odot}$ per component. However, as
in the thin-shell models, the warm-hot gas could harbor 10-150$\times$
more mass in such filaments \cite{refsom}. In either case (shells or
filaments), given the similarity of the cold and warm-hot absorption
lines (Fig.~\ref{navcompare}), the Ne~\textsc{viii}-bearing plasma
must be a transitional phase that links the colder and hotter material
and thus provides insights on the outflow physics. The
Ne~\textsc{viii}/N~\textsc{v} phase is not photoionized, so it must be
generated through interaction of the cold gas with a hotter ambient
medium analogous to X-ray emitting regions seen in nearby galaxies.
How this occurs is an open question; the absorbers could be material
cooling from the hot phase down to the cool gas, or the cool clouds
could have a hotter and more ionized surface that is evaporating.

Low-density plasma in the $T = 10^{5} - 10^{6}$ K range has been
effectively hidden from most outflow studies.  In principle, the
O~\textsc{vi} 1032,1038 \AA\ doublet can reveal such gas, but it is
unclear whether the O~\textsc{vi} arises in photoionized $10^{4}$ K
gas, hotter material at $\approx 10^{5.5}$ K, or both \cite{tripp08}.
The Ne~\textsc{viii} doublet avoids this ambiguity, and we have found
that this warm-hot matter is a substantial component in the mass
inventory of a galactic wind. Moreover, this wind has a large spatial
extent, and the mass carried away by the outflow will affect the
evolution of the galaxy. While earlier studies of poststarburst
outflows focused on Mg~\textsc{ii} and could not precisely constrain
the metallicity, hydrogen column, and mass, these studies do indicate
that poststarburst outflows are common: 22/35 of the poststarbursts in
\cite{tremonti09} showed outflowing Mg~{\sc ii} absorption with
maximum (radial) velocities of 500$-$2400 km s$^{-1}$, similar to the
absorption near 177\_9 (Figure~\ref{fig_llplot}), and 77\% and 100\%
of the poststarburst and AGN galaxies, respectively, in \cite{coil11}
drive outflows but with lower maximum velocities, which may be due to
selection of wind-driving galaxies in a later evolutionary stage.
With existing COS data, the effects of large-scale outflows on galaxy
evolution can be studied with the techniques presented here but with
larger samples \cite{tumlinson11}, with which it will be possible to statistically track how
outflows affect galaxies.

\vspace{0.5cm}
\noindent {\bf Supporting Online Material:} \\
Figs. S1 to S5 \\
Tables S1 to S2 \\
References (40-54) \\

\begin{table*}
\scriptsize
\centering
{\normalsize Table S1: Column Densities in the $z_{\rm abs}$ = 0.927 Absorption System of PG1206+459} \\
 \begin{tabular}{llcccc}
\hline \hline
 Species    & Fitted             & \multicolumn{4}{c}{log [Column Density (cm $^{-2}$)]} \\
  \         & Transitions (\AA ) & \multicolumn{4}{c}{(component velocity centroid, km s$^{-1}$)} \\
\hline
\multicolumn{6}{c}{Component Group A ($-450 < v < 0$ km s$^{-1}$)} \\
\hline
  \         &  \                 & (-317)         & (-247)         & (-162)         & (-88) \\
{\sc H i}   & Ly$\alpha$, Ly$\zeta$, Ly$\eta$   & 14.96$\pm$0.11 & 15.42$\pm$0.03 & 15.39$\pm$0.04 & 15.28$\pm$0.04 \\          
{\sc C ii}  & 687.05, 1334.53    & 13.65$\pm$0.03 & 13.83$\pm$0.03 & 13.80$\pm$0.05 & 13.51$\pm$ 0.04 \\                        
{\sc N ii}  & 644.63             & 12.71$\pm$0.25 & 13.50$\pm$0.08 & 13.09$\pm$0.13 & 13.01$\pm$ 0.14 \\                        
Mg~{\sc ii} & 2796.35, 2803.53   & 11.93$\pm$0.03 & 12.47$\pm$0.02 & 12.36$\pm$0.01 & 12.16$\pm$0.02 \\                         
Si~{\sc ii} & 1190.42,1193.29,1260.42,1526.71 & 12.25$\pm$0.14 & 12.24$\pm$0.17 & 12.52$\pm$0.08 & 12.43$\pm$0.11 \\            
{\sc N iii} & 685.00, 763.34     & 14.50$\pm$0.02 & 14.42$\pm$0.03 & 14.20$\pm$0.04 & 14.02$\pm$0.05 \\                         
{\sc O iii} & 702.33             & $\geq$14.65    & $\geq$14.65    & $\geq$14.52    & $\geq$14.55 \\ 
{\sc S iii} & 698.73, 724.29     & 13.17$\pm$0.03 & 13.21$\pm$0.03 & 13.19$\pm$0.03 & 12.88$\pm$0.06 \\                         
{\sc N iv}  & 765.15             & $\geq$14.24    & $\geq$14.37    & $\geq$14.14    & $\geq$14.06 \\ 
{\sc O iv}  & 787.71             & $\geq$14.91    & $\geq$14.88    & $\geq$14.82    & $\geq$14.86 \\ 
Si~{\sc iv} & 1393.76, 1402.77   & 13.00$\pm$0.04 & 12.94$\pm$0.05 & 13.05$\pm$0.09 & 13.16$\pm$0.04 \\                         
{\sc S iv}  & 657.34             & 13.20$\pm$0.03 & 13.25$\pm$0.03 & (blended)      & (blended)      \\                         
{\sc N v}   & 1238.82, 1242.80   & 14.30$\pm$0.04 & 14.30$\pm$0.05 & 13.40$\pm$0.18 & 13.96$\pm$0.08 \\                         
{\sc S v}   & 786.48             & 12.88$\pm$0.03 & 12.84$\pm$0.03 & 12.67$\pm$0.05 & 12.31$\pm$0.08 \\                         
Ne~{\sc viii}& 770.41, 780.32    & 13.71$\pm$0.29 & 14.04$\pm$0.08 & (not detected) & 14.07$\pm$0.04 \\                         
Mg~{\sc x}  & 624.95             & \multicolumn{2}{c}{$\longleftarrow \ \ \ <14.04 \ \ \ \longrightarrow $} & \multicolumn{2}{c}{$\longleftarrow \ \ \ <13.94 \ \ \ \longrightarrow $} \\ 
\hline
\multicolumn{6}{c}{Component Group B ($0 < v < 300$ km s$^{-1}$)} \\
\hline
  \         & \                  & (65)          & (+108)           & (+159)          &  \\
{\sc H i}   & Ly$\alpha$, Ly$\zeta$, Ly$\eta$, Ly$\nu$    & 16.28$\pm$0.13 & 16.76$\pm$0.08 & 16.32$\pm$0.07 & \ \\ 
{\sc C ii}  & 687.05, 1334.53    & $\geq$14.10    & $\geq$14.36    & 13.88$\pm$0.07 & \ \\              
{\sc N ii}  & 644.63             & 14.01$\pm$0.04 & 14.04$\pm$0.05 & 13.62$\pm$0.12 & \ \\                          
Mg~{\sc ii} & 2796.35, 2803.53   & 13.39$\pm$0.03 & 13.36$\pm$0.01 & 12.84$\pm$0.01 & \ \\                          
Si~{\sc ii} & 1190.42,1193.29,1260.42,1526.71 & 13.50$\pm$0.12 & 13.39$\pm$0.04 & 13.06$\pm$0.09 & \\               
{\sc N iii} & 763.34             & 14.64$\pm$0.04 & 14.54$\pm$0.13 & 14.65$\pm$0.12 &  \\                           
{\sc O iii} & 702.33             & (saturated)          & (saturated)          & (saturated) &     \\               
{\sc S iii} & 698.73, 724.29     & 13.62$\pm$0.04 & 13.58$\pm$0.04 & 13.66$\pm$0.03 &  \\                           
{\sc N iv}  & 765.15             & (saturated)    & (saturated)    & (saturated)    &  \\                           
{\sc O iv}  & 608.40, 787.71     & (saturated)    & (saturated)    & (saturated) \\                                 
Si~{\sc iv} & 1393.76, 1402.77   & 13.32$\pm$0.03 & 13.28$\pm$0.04 & 13.47$\pm$0.03 & \\                            
{\sc S iv}  & 657.34             & 13.47$\pm$0.04 & 13.13$\pm$0.09 & 13.66$\pm$0.04 &  \\                           
{\sc N v}   & 1238.82, 1242.80   & 13.93$\pm$0.06 & 13.94$\pm$0.10 & 13.84$\pm$0.11 & \\                            
{\sc S v}   & 786.48             & 12.93$\pm$0.04 & (blended)      & (blended)      &  \\                           
Ne~{\sc viii}& 770.41, 780.32    & \multicolumn{2}{c}{$ \longleftarrow \ \ 14.53 \pm 0.04 \ \ \longrightarrow$ } & 14.21$\pm$0.05 \\ 
Mg~{\sc x}  & 624.95             & \multicolumn{3}{c}{$\longleftarrow \ \ \ <14.20 \ \ \ \longrightarrow $} &  \\   
\hline
\multicolumn{6}{c}{Component Group C ($1000 < v < 1250$ km s$^{-1}$)} \\
\hline
  \         & \                               & (+1084)         & (+1131)        & \              &  \\
{\sc H i}   & Ly$\alpha$, Ly$\zeta$, Ly$\iota$, Ly$\xi$   & 15.02$\pm$0.16 & 15.93$\pm$0.03 & \ & \\ 
{\sc C ii}  & 687.05                          & 12.92$\pm$0.14 & 13.58$\pm$0.04 & \ & \ \\           
{\sc N ii}  & 644.63                          & (blended)      & (blended)      & \ & \\             
Mg~{\sc ii} & 2796.35, 2803.53                & $< 11.37$      & 12.10$\pm$0.04 & \ & \\             
Si~{\sc ii} & 1190.42,1193.29,1260.42,1526.71 & $< 12.37$      & 12.53$\pm$0.12 & \ & \\             
{\sc N iii} & 685.00, 763.34                  & (blended)      & (blended)      & \ & \\             
{\sc O iii} & 702.33                          & 13.48$\pm$0.09 & $\geq$14.64    & \ &  \\      
{\sc S iii} & 698.73, 724.29                  & $< 12.33$      & 13.55$\pm$0.03 & \ & \\             
{\sc N iv}  & 765.15                          & 12.53$\pm$0.12 & 13.59$\pm$0.02 & \ &  \\            
{\sc O iv}  & 608.40, 787.71                  & 13.95$\pm$0.05 & $\geq$15.02    & \ &  \\      
Si~{\sc iv} & 1393.76, 1402.77                & (not detected) & 13.63$\pm$0.05 & \ & \\             
{\sc S iv}  & 657.34                          & 12.18$\pm$0.35 & 13.30$\pm$0.03 & \ &  \\            
{\sc N v}   & 1238.82, 1242.80                & $< 13.49$      & $< 13.56$      & \ & \\             
{\sc S v}   & 786.48                          & 11.87$\pm$0.32 & 13.03$\pm$0.03 & \ &  \\            
Ne~{\sc viii}& 770.41, 780.32                  & 13.30$\pm$0.27 & 13.78$\pm$0.09 & \ & \\ 
Mg~{\sc x}  & 624.95                          & \multicolumn{2}{c}{$\longleftarrow \ \ \ <14.08 \ \ \ \longrightarrow $} & \ & \\ 
\hline
\end{tabular}
\end{table*}

\paragraph*{Data reduction and column density measurements.}
We have reduced the Cosmic Origins Spectrograph observations of
PG1206+459 as described in \cite{meiring11}, including careful
alignment of individual exposures (which are intentionally offset on
the detector to reduce the effects of fixed-pattern noise),
flatfielding, and direct determination of uncertainties from the source
and background counts (COS has a photon-counting detector).  The default COS data are highly oversampled,
so we binned the spectra to Nyquist sampling of the 15 km s$^{-1}$
spectral resolution of the instrument.

Our column-density measurements are summarized in Table~S1. A
unique aspect of the HST/COS data presented in this paper is the
coverage of lines in the rest-frame far-ultraviolet (FUV) at
wavelengths between 600 \AA\ and 912 \AA . These lines cannot be
observed in the Milky Way because interstellar hydrogen completely
absorbs light at these wavelengths from any sources beyond a small
region around the Sun. This is unfortunate because the FUV is rich in
spectral diagnostics of physical conditions and gas composition
(\textit{23}). Indeed, these lines are rarely observed in any
astrophysical context, although they are powerful tools for solar
physics. To access these diagnostics, we use quasars to search for
absorption lines at redshifts $z >$ 0.5 so that the FUV transitions
are redshifted into the standard HST bandpass ($\lambda _{\rm HST} >
1150$ \AA ).  For example, $z_{\rm abs} >$ 0.5 redshifts the Ne~{\sc
  viii} 770.41,780.32 \AA\ doublet into the wavelength range
observable with HST.  In principle, these lines can be studied from
the ground for sufficiently high $z_{\rm abs}$ systems, but in
practice this is extremely difficult because the ``Ly$\alpha$ forest''
imprints hundreds of closely spaced absorption lines on the spectrum
at such high redshifts and makes detailed analysis of FUV metal lines
very difficult.  More importantly, high-redshift QSO sightlines have a
high probability of encountering an optically thick Lyman limit
absorber that completely absorbs the QSO flux in the rest-frame FUV
range.  Consequently, we have elected to study FUV absorption lines at
$z_{\rm abs} <$ 1.5, which cannot be observed from the ground but are
in a redshift range where the Ly$\alpha$ forest is much thinner,
blending is less common, and optically thick LL systems are rare (even
in this redshift range, some blending with interloping lines from
other redshifts occurs, as can be seen from Figure~2).  Accordingly,
we have targeted QSOs at $z_{\rm QSO} \approx 1 - 1.5$ in order to
have a large total redshift path over which to search for the
Ne~\textsc{viii} doublet and the other FUV lines.  Spectrographs on
the HST before COS and on the Far Ultraviolet Spectroscopic Explorer
generally could not target QSOs at $z \gg 0.5$ because they are mostly
too faint to observe with those instruments in practical exposure
times.  Consequently, the Ne~\textsc{viii} doublet has been only
rarely detected ({\it 27}), and some of the lines in Table~S1 have
never been seen in interstellar or intergalactic contexts.

\begin{figure}
\includegraphics[width=9.0cm, angle=0]{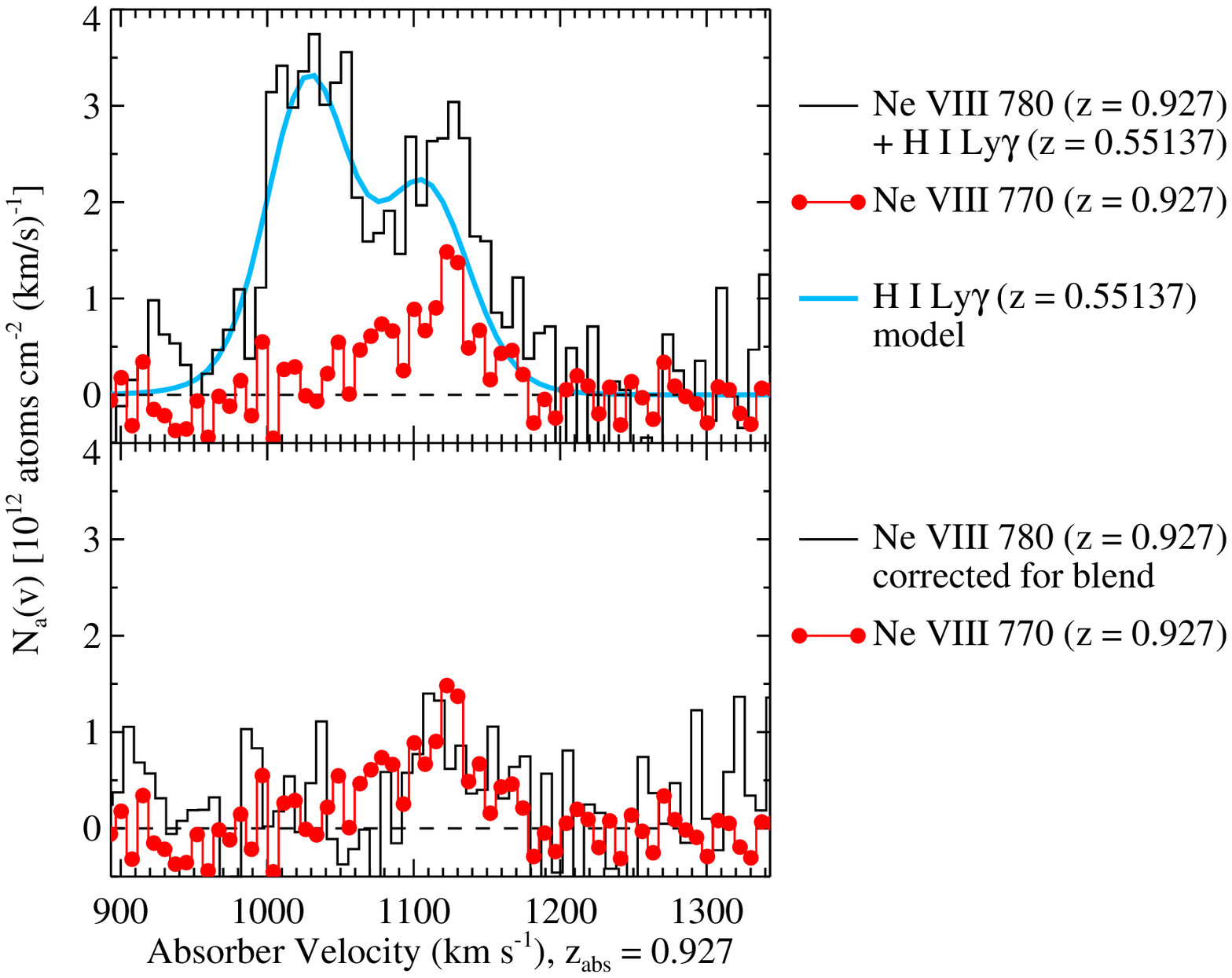}
{Figure S1: (Top) The apparent column density profiles
  (\textit{28}) of Ne~{\sc viii} 780 \AA\ (black lines)
  and Ne~{\sc viii} 770 \AA\ (red lines with dots) before
  the 780 \AA\ line is corrected for a blend with {\sc H~i} Ly$\gamma$
  at $z_{\rm abs}$ = 0.55137.  The smooth blue line shows a model of
  the Ly$\gamma$ line, based on the well-detected {\sc H~i} Ly$\beta$
  and Ly$\epsilon$ lines at the same redshift (Ly$\delta$ is lost in a
  strong blend).  (Bottom) By correcting the data with the
  H~\textsc{i} Ly$\gamma$ model, the {\sc H~i} blend is effectively
  removed, and the residual profile of Ne~{\sc viii} 780
  \AA\ is found to be in excellent agreement with 770 \AA .}
\end{figure}

To overcome the occasional blending problem, we use two strategies to
measure column densities.  First, we use as many transitions as
possible to constrain a given species by simultaneously fitting Voigt
profiles to all of the lines.  For example, we simultaneously fit the
{\sc S~iii} lines at 698.73 and 724.29 \AA\ (in principle, we could
also fit the {\sc S~iii} line at 677.75 \AA , but this line is
redshifted into a region that is badly affected by {\sc O~i} emission
from the Earth's exosphere, and at the time of this writing, the COS
data reduction software could not exclude that contamination).  In
velocity ranges where one transition is affected by a strong blend,
that velocity range is masked out in the problematic line but is still
constrained by the other transitions used in the fit.  Second, in some
cases we can estimate the strength of the blended interloper and
divide it out of the profile of interest.  An example of this
technique is shown in Figure~S1.  The upper panel of this figure
compares the apparent column density profiles ({\it 28}) of the
Ne~{\sc viii} 770.41,780.32 \AA\ lines in the component at $v = +1131$
km s$^{-1}$.  In this component, the Ne~{\sc viii} 780.32 \AA\ line is
blended with an {\sc H~i} Ly$\gamma$ line at $z_{\rm abs} =
0.55137$. Fortunately, the {\sc H~i} Ly$\beta$ and Ly$\epsilon$ lines
are also clearly detected at $z_{\rm abs} = 0.55137$, and by fitting
the Ly$\beta$ and Ly$\epsilon$ transitions, we can predict the
strength of the Ly$\gamma$ line and remove it from the blend with the
Ne~{\sc viii} 780.32 \AA\ profile.  The Ly$\gamma$ line modeled in
this way is shown with a smooth blue line in the upper panel of
Figure~S1, and the lower panel of Figure~S1 shows the Ne~{\sc viii}
780.32 \AA\ profile after the correction for the Ly$\gamma$ blend has
been applied. In the lower panel the two Ne~{\sc viii} apparent column
density profiles are in excellent agreement.  We have used similar
modeling to correct for the {\sc H~i} Ly$\beta$ line at $z_{\rm abs}$
= 0.46435, which is somewhat blended with the Ne~{\sc viii} 780.32
\AA\ line in the component at $v = -317$ km s$^{-1}$.

With this approach, we have measured the column densities of a wide
array of low- , intermediate- , and high-ionization metals in the
PG1206+459 absorption system at $z_{\rm abs}$ = 0.927, and our column
density measurements are listed in Table~S1. For convenience in
comparisons with previous work (\textit{24}), we divide the components
into the three groups they defined (groups A, B, and C).  We fit nine
components, the maximum number we can effectively constrain with free
fits to the COS data (four components in group A, three in group B,
and two in group C). In some cases, the lines are so strong that
saturation precludes a useful column density measurement, and in these
cases we simply indicate that the component is ``saturated'' in
Table~S1.  In other cases, the lines are strong enough to be
moderately saturated but at a level that can be inferred from the
profile shapes, especially when multiple lines of varying strength are
fitted simultaneously.  In these cases we report the column density
indicated by profile fitting in Table~S1, but we indicate the line as
a lower limit because the profile fitting may still suffer significant
systematic uncertainty due to saturation.

\paragraph*{Properties of the affiliated galaxy.} 

\begin{figure}
\includegraphics[width=8.00cm, angle=0]{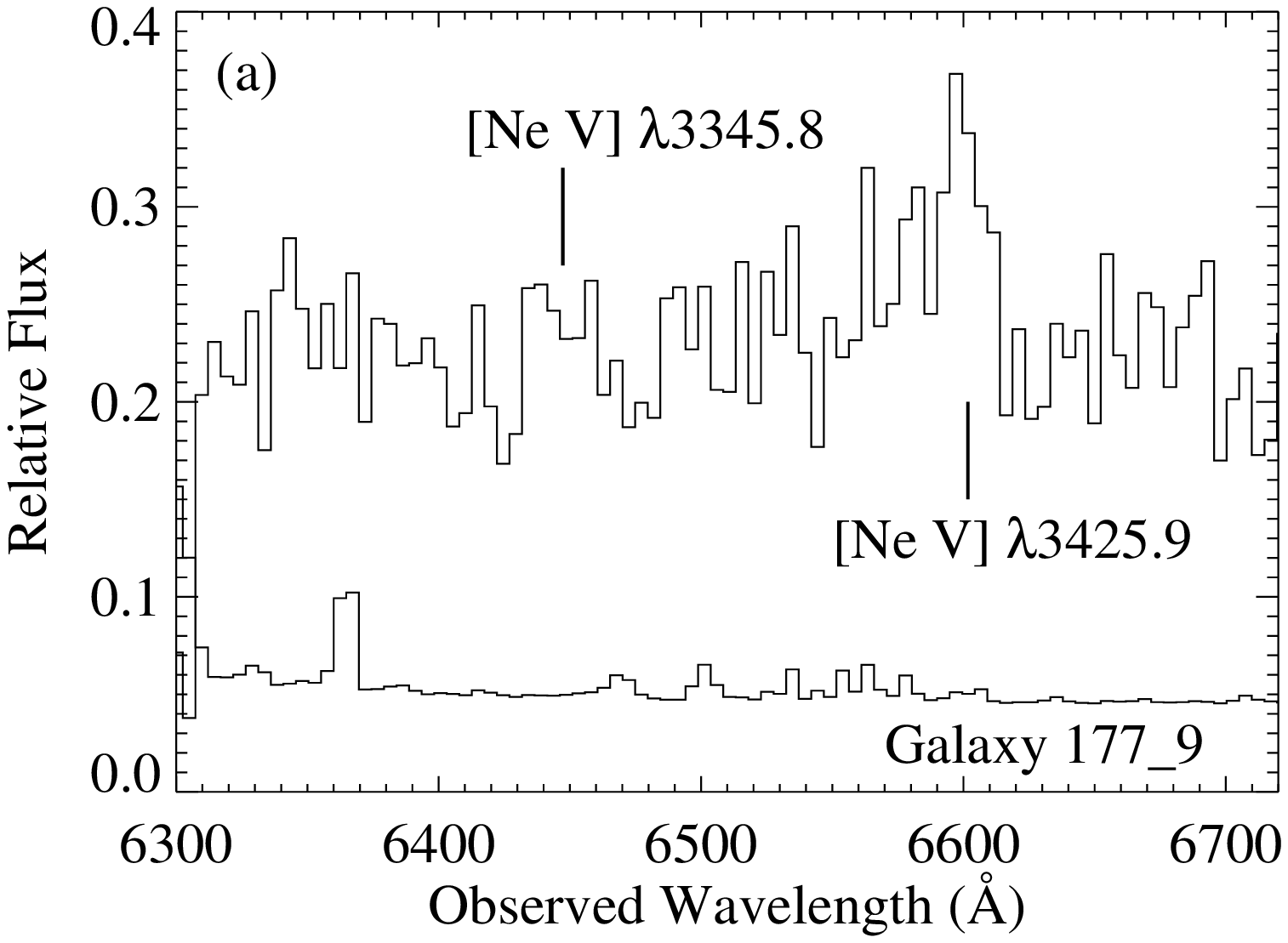}
\includegraphics[width=8.00cm, angle=0]{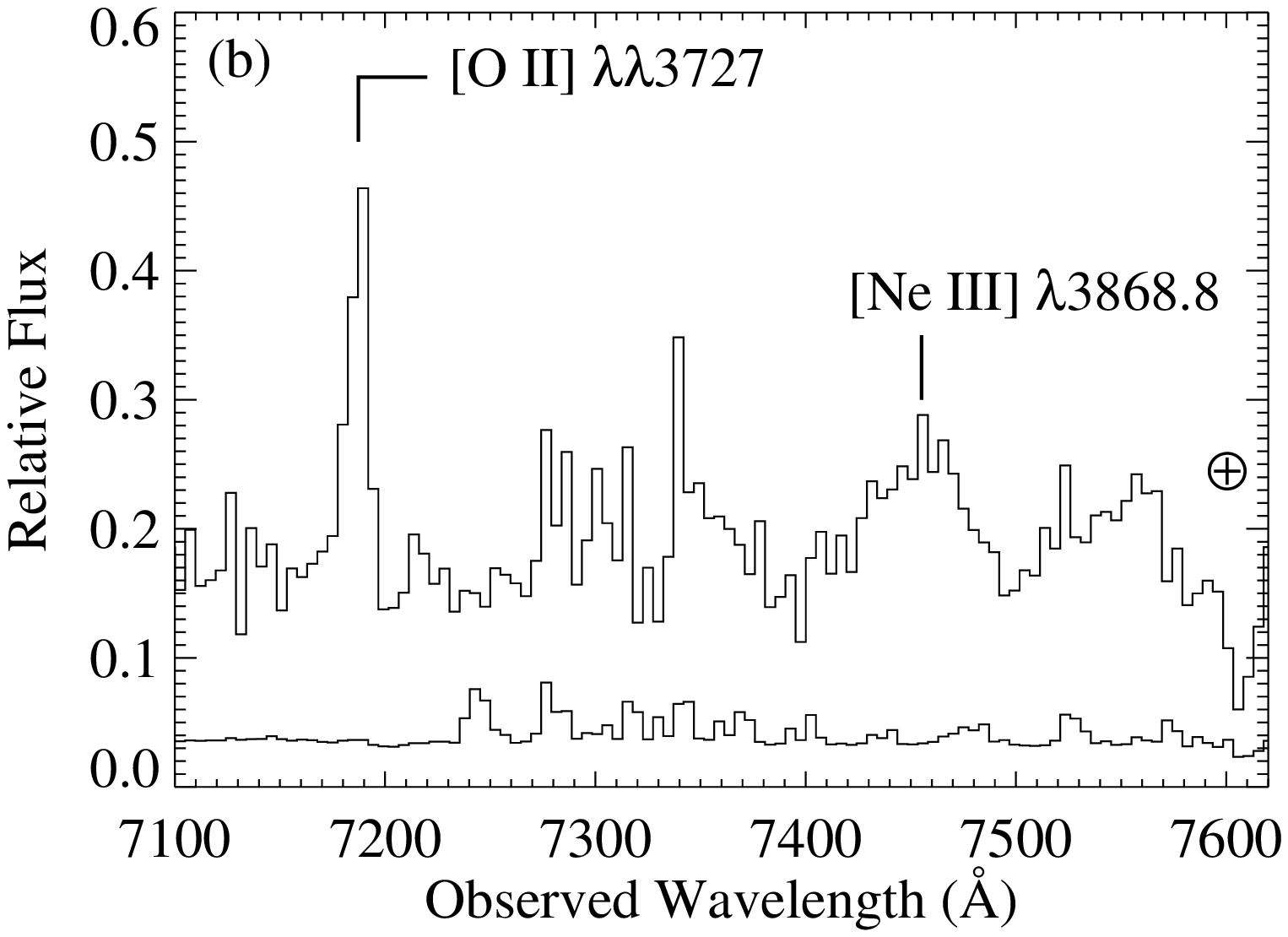}
{Figure S2: Expanded views of portions of the MMT spectrum of galaxy
  177\_9 from Figure~4 including (a) the region of
  the spectrum near the [Ne~\textsc{v}] lines at 3345.8 and 3425.9 \AA
  , and (b) the region near the [O~\textsc{ii}] emission at 3727
  \AA\ and [Ne~\textsc{iii}] emission at 3868.8 \AA .  As in
  Figure~4, the upper histogram shows the flux from
  the galaxy and the lower histogram shows the 1$\sigma$ flux
  uncertainty.}
\end{figure}

The MMT spectrum of galaxy 177\_9 (Figure~4), shows
highly significant [O~\textsc{ii}] emission along with a 4$\sigma$
emission feature that we identify as the [Ne~\textsc{v}] 3426
\AA\ transition.  This [Ne~\textsc{v}] 3426 emission is often a
signature of an active galactic nucleus \cite{shields10}; however, it
is expected to be accompanied by weaker [Ne~\textsc{v}] and
[Ne~\textsc{iii}] emission features at 3346 and 3869
\AA\ \cite{shields10}. These weaker emission features are not
immediately obvious in Figure~4, which calls the
[Ne~\textsc{v}] 3426 identification into question.  However, as we
show in Figure~S2, the weaker neon emission lines are confused by
ambiguous continuum placement as well as the intrinsic noisiness of
the spectrum.  A weak feature is evident at the expected wavelength of
[Ne~\textsc{v}] $\lambda$3346 (Figure~S2a).  The significance of this
feature depends on where the galaxy continuum is drawn in, but it is
expected to be a factor of $\sim$3 weaker than [Ne~\textsc{v}]
$\lambda$3426 and thus is consistent with the stronger [Ne~\textsc{v}]
line.  The [Ne~\textsc{iii}] $\lambda$3869 line (Figure~S2b) is nested
between absorption lines and is close to the 4000 \AA\ break as well
as a strong telluric absorption feature.  The complexity of this
spectral region, combined with noise, makes the continuum placement
highly uncertain, and a [Ne~\textsc{iii}] $\lambda$3869 feature with
the expected strength could easily be present.  Given the good
agreement of the [Ne~\textsc{v}] $\lambda$3426 and [O~\textsc{ii}]
$\lambda$3727 features, we conclude that the [Ne~\textsc{v}]
$\lambda$3426 identification is reliable.  In turn, this implies that
the [O~\textsc{ii}] $\lambda$3727 emission might not be a suitable
indicator of the star formation rate in 177\_9 because the
[O~\textsc{ii}] could be predominantly due to the AGN emission-line
region \cite{abel08,yan06}.

While it is interesting that the spectral features, high luminosity,
and blue rest-frame color of galaxy 177\_9 are similar to those of
poststarburst galaxies ({\it 11,37}), we caution that its
color and spectral features are noisy, and it is necessary to reduce
these uncertainties to reliably classify the galaxy.  The galaxy color
is uncertain because it is derived from the Sloan Digital Sky Survey
(SDSS) photometry of this object, and the galaxy is close to the SDSS
detection limit. Unfortunately, the LBT images were not obtained in
conditions suitable for photometry.  For comparison with the DEEP2
galaxy color-magnitude diagram (Figure~4), we have
transformed the SDSS photometry of 177\_9 into a rest-frame U-B color
using the K-correction and methods of (\textit{29}), and this
introduces some systematic uncertainty.  However, the poststarburst
galaxies from (\textit{11}) shown in Figure~4
were transformed from SDSS filters to rest-frame U-B in an identical
fashion, and the offset of these poststarburst galaxies from the
``blue cloud'' of ordinary galaxies is similar to the offset seen in
SDSS filters directly (\textit{37}). Deeper photometry of 177\_9
would help to elucidate its nature.

We also note that another bright galaxy, apart from 177\_9, is evident
close to the PG1206+459 sight line (the yellow object northwest of the
QSO in Fig.~4). We obtained an MMT spectrum of this
galaxy, but unfortunately the spectrum does not reliably constrain its
redshift.  However, the absence of features in the wavelength range of
the MMT spectrum, the angular extent of this galaxy, and its color
suggest that it has $z \ll$ 0.927.  This conclusion is corroborated by
the photometric redshifts of this object from the SDSS database, which
range from $z_{\rm phot} = 0.21$ \cite{csabai03} to $z_{\rm phot} =
0.43$ \cite{oyaizu08}.

\paragraph*{Ionization, metallicity, and mass: cool phase.} 
We obtain accurate {\sc H~i} measurements from the Lyman series lines
(Figs.~1$-$2). The {\sc H~i} column
densities are not high enough to prevent photoionization by the UV
background light. Consequently, to estimate the total mass we apply
standard photoionization models to account for the ionized hydrogen
({\sc H~ii}) using CLOUDY, v8.0 (\textit{33}). We assume the gas is
photoionized by the diffuse UV background from quasars \cite{haardt96}
and require the models to match the observed $N$({\sc H~i}), $N$({\sc
  S~iii}), and $N$({\sc S~iv}) on a component-by-component basis. We
prefer {\sc S~iii} and {\sc S~iv} because sulfur does not deplete onto
dust and the lines are weak enough so that they are not confused by
unresolved saturation. Some components are blended with unrelated
lines from other redshifts, but we account for blends (and in some
cases remove them) by employing multiple transitions (see above). We
characterize the models by the ionization parameter ($\equiv$ ionizing
photon density/particle density), which is adjusted to fit the {\sc
  S~iii}/{\sc S~iv} ratio, and then the metallicity is adjusted to fit
the observed column densities with $N$(H~\textsc{i}) fixed to the
observed value.  We have modeled the components at $v = -317,-247,65,$
and 1131 km s$^{-1}$ (the components with good {\sc S~iii}/{\sc S~iv}
measurements); examples of the models and results are shown in
Figure~S3. We do not expect these models to fit perfectly -- there are
many uncertain model parameters such as the shape and intensity of the
UV background, the relative abundance patterns (we assume the solar
pattern), and confusion due to dust depletion. Given these caveats,
the models (Figure~S3) compare reasonably well to the observations --
with the ionization parameter fixed by {\sc S~iii}/{\sc S~iv}, most of
the remaining low- and intermediate-ionization stages agree with the
models to within 0.3 -- 0.5 dex.  We obtain several robust results:
First, while the metallicity is uncertain by $\approx 0.3$ dex due to
radiation field uncertainties, the gas must be highly enriched
(Table~S2). Second, the observed Ne~{\sc viii}, {\sc N~v}, and {\sc
  S~v} columns are significantly higher (in some cases, by orders of
magnitude) than predicted by the photoionization models. These species
must arise in a separate, higher ionization phase. Third, the models
require high total hydrogen ({\sc H~i} + {\sc H~ii}) columns even with
high metallicities, and this in turn implies large masses.

\begin{figure}
\includegraphics[width=11.5cm, angle=0]{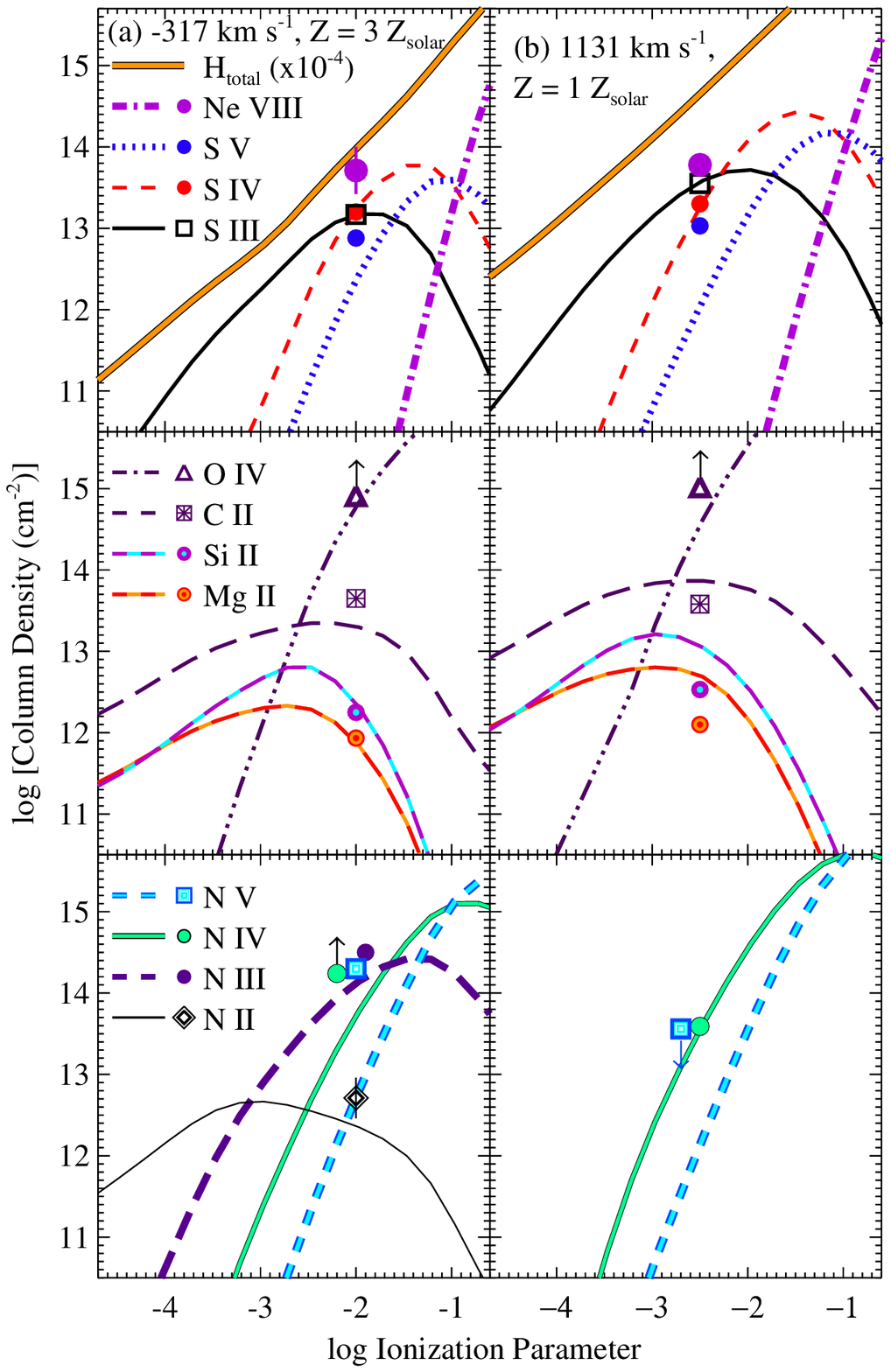}
{Figure S3: Comparison of the observed column densities in
  two of the absorber components to predictions from a CLOUDY
  (\textit{33}) photoionization model. The columns show results for
  the components at (a) $-317$ km s$^{-1}$, and (b) +1131 km s$^{-1}$.
  In each panel, the column densities predicted by the model are shown
  with smooth curves and the observed columns are plotted with
  discrete symbols at the ionization parameter that provides the best
  fit to the observed {\sc S~iii} and {\sc S~iv} column
  densities. Some species (e.g., N~\textsc{iv}, N~\textsc{v},
  S~\textsc{v}, Ne~\textsc{viii}) do not agree with the models due to
  contributions from the hot-gas phase (Figure~S5).  The
  species corresponding to the curves and discrete symbols are
  indicated by the legends in column (a). Column density uncertainties
  are usually smaller than the symbol sizes.  The
  underabundance of Mg~\textsc{ii} and Si~\textsc{ii} in (b) could be
  due to dust, which depletes these species, but these ions are
  sensitive to the shape of the ionizing flux field.}
\end{figure}

We can obtain a rough estimate of the mass ($M$) in the outflowing
components using a standard thin-shell model \cite{rupke05b}: $M = \mu
m_{p} N_{\bot}({\rm H})_{\rm total} \Omega r^{2}$, where $r$ is the
shell radius ($\geq$ impact parameter = 68 kpc), $N_{\bot}({\rm
  H})_{\rm total}$ is the total H column perpendicular to the shell
surface, $\Omega$ is the shell solid angle, $m_{p}$ is the proton
mass, and $\mu = 1.3$ accounts for the additional mass of helium.  To
calculate $\Omega$, we assume the outflow full opening angle $\theta$
ranges between 45 and 135$^{\circ}$ as observed in nearby starburst
and AGN outflows (\textit{6}).  Unlike many previous studies, we
are not observing a sight line directly toward the galaxy that
launches the outflow, so me must convert our observed total hydrogen
column densities (along the line of sight) into $N_{\bot}({\rm
  H})_{\rm total}$.  To make this conversion, we adopt the simplified
geometry shown in Figure~S4. We detect absorption at both negative and
positive velocities with respect to the galaxy
(Figure~2), and despite the large velocity separation,
the positive- and negative-velocity gas has similar properties.  This
can be understood if we are detecting the front and back sides of a
biconical outflow, as illustrated in Figure~S4, but this requires that
the sight line is roughly parallel to the outflow axis (or else only
the front or back side shell would be detected, but not both).  In
this situation, if the sight line to the background QSO is at large
enough impact parameter ($\rho$), it will not penetrate the shells and
no absorption will be detected (position A in Figure~S4).  At the
maximum impact parameter that intercepts the shells (position B), we
see that $r = \rho / sin (\theta /2)$ and $N_{\bot}$(H) = $N$(H)$_{\rm
  total}$\ cos ($\theta$/2).  As the sight line is moved to smaller
impact parameters (e.g., position C), $r$ will increase and
$N$(H)$_{\rm total}$ $\longrightarrow N_{\bot}$(H), both of which will
increase the mass compared the value calculated assuming the sight
line is at position B.  Thus, by calculating the mass assuming the
sight line is near position B, we derive a lower limit.  With these
assumptions, we obtain the $N_{\bot}$(H) values and masses summarized
in Table~S2 for the cool (low-ionization) phase.

\begin{figure}
\includegraphics[width=9.0cm, angle=90]{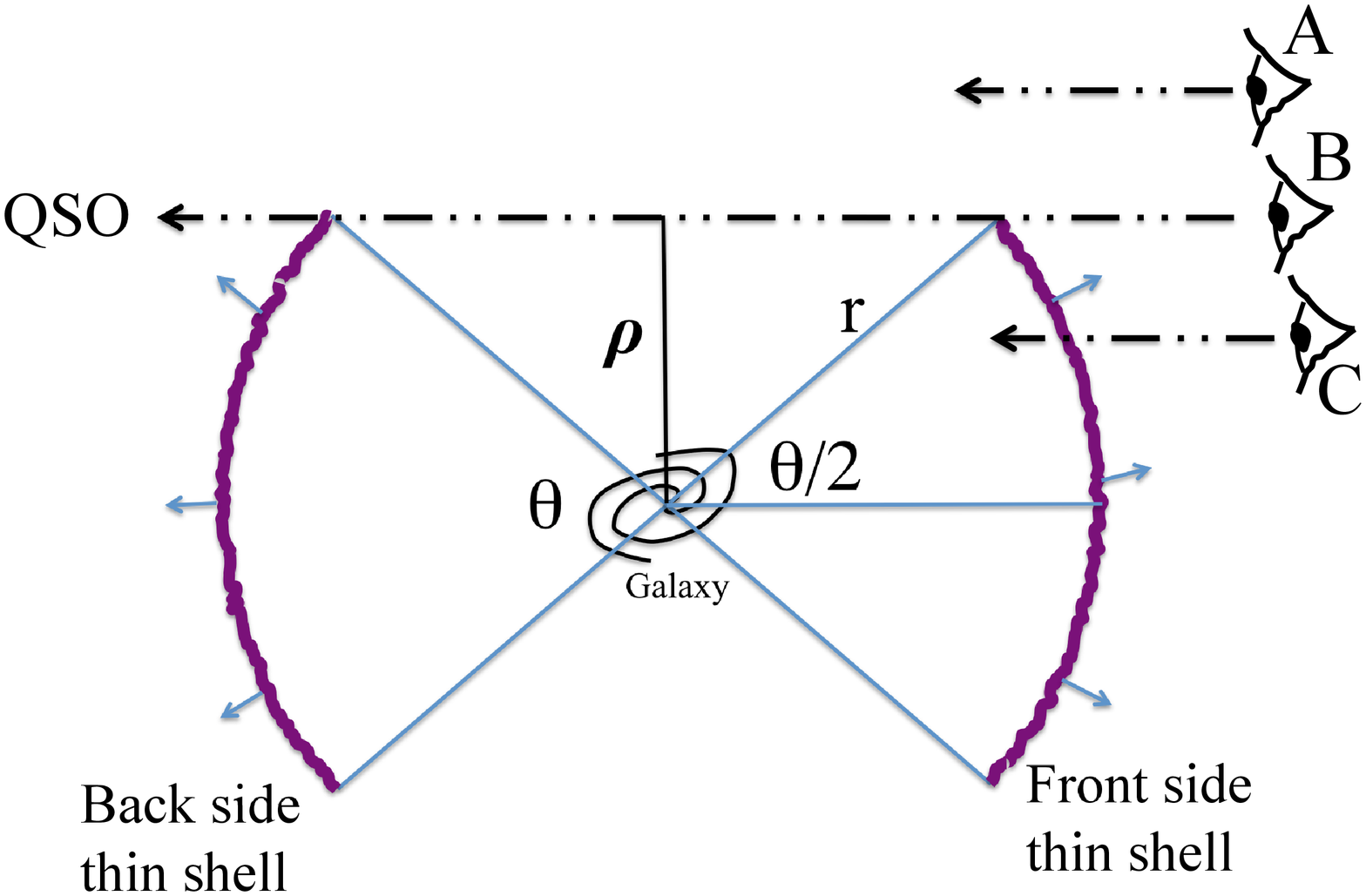}
{Figure S4: Schematic sketch of the simplified geometry assumed to
  roughly estimate shell mass.  We assume that the sight line to the
  background QSO is roughly parallel to the outflow axis so that the
  front and back sides of the biconical outflow are intercepted by the
  sight line.  We assume that the shells are at the cone caps, so an
  observer at position A would not detect the outflow because that
  sight line does not intercept the bicone.  Position B represents the
  maximum impact parameter that pierces the shell, and at this
  position, $N_{\bot}$(H) = $N$(H)$_{\rm total}$\ cos ($\theta$/2),
  where $\theta$ is the full opening angle. As the observer moves from
  position B toward position C and toward viewing the galaxy directly,
  $N$(H)$_{\rm total}$ $\longrightarrow N_{\bot}$(H), and $r$
  increases, both of which increase the calculated mass in the shell.
  Therefore, the mass calculated assuming position B provides a lower
  limit on the true mass.}
\end{figure}

\begin{table*}
 \scriptsize
 \centering
{\normalsize Table S2: Metallicities (Elemental Abundances) and Thin-Shell Masses
   of Individual Components in the Absorption Line
   System}
 \begin{tabular}{llccccccc}
 \hline \hline
 \ & \multicolumn{4}{c}{\underline{ \ \ \ \ \ \ \ \ \ \ \ \ \ \ \ \ \ \ \ \ \ \ \ \ \ \ \ Cool Phase \ \ \ \ \ \ \ \ \ \ \ \ \ \ \ \ \ \ \ \ \ \ \ \ \ \ \ }} & \multicolumn{4}{c}{\underline{ \ \ \ \ \ \ \ \ \ \ \ \ \ \ \ \ \ \ \ \ \ \ \ \ \ \ \ \ \ \ Hot Phase \ \ \ \ \ \ \ \ \ \ \ \ \ \ \ \ \ \ \ \ \ \ \ \ \ \ \ \ \ \ }} \\
Component     &     \               &     \               & \multicolumn{2}{c}{(Thin-shell model)}    &         \              & \multicolumn{2}{c}{(Thin-shell model)} & Gas \\
Velocity      & Metallicity         & log $N$(H)$_{\rm total}$ & log $N_{\bot}$(H)$_{\rm total}$ & Mass            & log $N$(H)$_{\rm total}$ & log $N_{\bot}$(H)$_{\rm total}$ & Mass          & Temperature     \\
(km s$^{-1}$) &     \               &     \               & \  & ($10^{8} M_{\odot}$)      & \ & \ & ($10^{8} M_{\odot}$) &  ($10^{5}$ K) \\ \hline
$-317$        & 3 $Z_{\rm solar}$   & 18.0 & 17.6$-$18.0         & 0.6$-$1     & 19.2$-$19.8          & 18.8$-$19.8           & 10$-$80   & $2.2 - 3.8$  \\
$-247$        & 1 $Z_{\rm solar}$   & 18.5 & 18.1$-$18.5         & 2$-$4       & 20.1$-$20.4          & $19.7 - 20.3$         & $80-300$  & $3.3 - 4.0$ \\
$+65$         & 0.5 $Z_{\rm solar}$ & 19.1 & 18.7$-$19.1         & 8$-$14      & 20.4$-$20.5          & $19.9 - 20.5$         & $100-400$ & $4.2 - 4.5$ \\
$+1131$       & 1 $Z_{\rm solar}$   & 18.6 & 18.2$-$18.6         & 3$-$5       & $\cdots$             & $\cdots$              & $\cdots$  & $\cdots$      \\ \hline
 \end{tabular}
\end{table*}

Of course, this model is highly simplified compared to the complex
situation that is likely in real outflows.  For example, the schematic
sight lines in Figure~S4 could be tilted toward the galaxy somewhat,
and front- and back-side absorption would still be detected.  However,
the sight line must have an impact parameter of 68 kpc, so as it is
tilted toward the galaxy, the radius becomes larger and $N$(H)$_{\rm
  total}$ $\longrightarrow N_{\bot}$(H), so our mass calculations are
still lower limits in this situation.  More importantly, the geometry
of the outflowing material could be completely different from thin
shells.  If the absorption arises in a context similar to the multiple
bubbles and filaments seen near some AGN in the nearby universe
\cite{blanton03,fabian11} or the multifilament outflows extending away
from the disks of nearby starburst galaxies (\textit{6}), then the
geometry is much more complex, although the absorption could still
originate in a thin shell or filament.  In filaments, the mass could
be substantially lower.  For example, the optical filaments of NGC1275
are long and skinny with dimensions of 0.07 kpc by 6 kpc (\textit{35})
and thus have a cold-gas mass $\approx 10^{6} M_{\odot}$.  However, as
discussed below, the total column density and mass in the
Ne~\textsc{viii} phase is substantially higher than the cold-gas mass.
The network of optical filaments centered on NGC1275 extend up to
$\approx 70$ kpc from the galaxy \cite{conselice01}, so the scale of
NGC1275 emission is similar to the scale of the absorption near galaxy
177\_9.  The complex velocity distribution of the absorption lines
near 177\_9 (Figure~1 - 2) -- with more components at negative
velocities than positive velocities -- could arise naturally in a set
of bubbles, shells, and filaments like those surrounding NGC1275.  We
note that magnetic fields have been suggested to stabilize the skinny
filaments of NGC1275 (\textit{35}) as well as many other central
cluster galaxies \cite{mcdonald10} and starburst galaxies
(\textit{6}).  It would be interesting to consider whether
magnetically confined gas could prevent the low- and high-ionization
gas from developing different kinematics and thus explain the striking
similarity and alignment of the low- and high-ionization lines in the
PG1206+459 absorber (Figure~3).

At this time, many configurations and models remain viable for the
absorption affiliated with 177\_9, so the thin-shell masses in
Table~S2 should be viewed as order-of-magnitude
calculations, but it is interesting to note that even though there are
large uncertainties, the absorbing gas is likely to contain substantial mass.
It is also possible to consider thick shells and clumping of the gas
within the shells \cite{rupke05b}, but more detailed observational
constraints must be obtained for such models.

\paragraph*{Ionization and mass: hot phase.}

\begin{figure}
\includegraphics[width=11.5cm, angle=0]{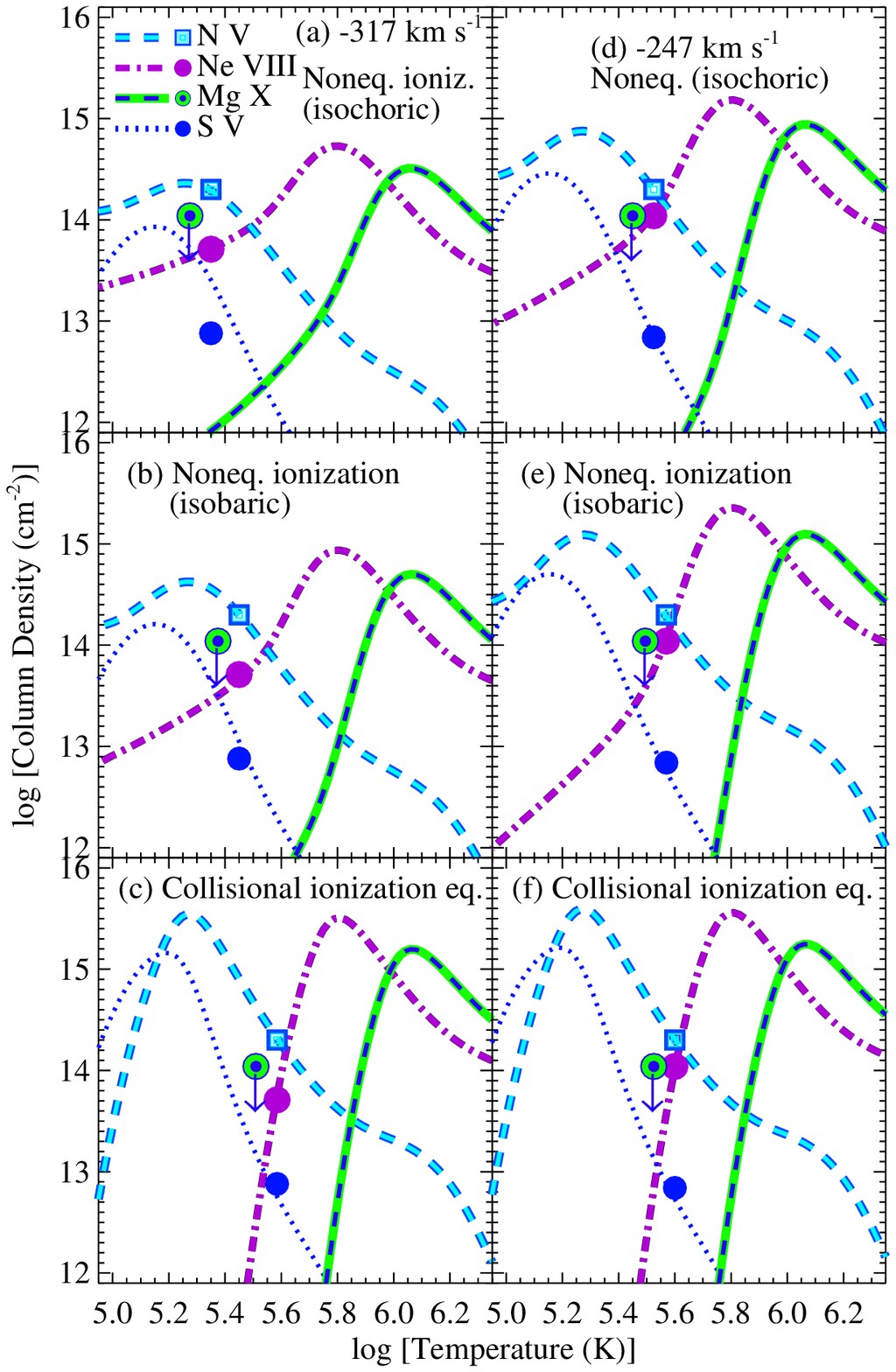}
{Figure S5: (a) Comparison of the observed N~\textsc{v},
  S~\textsc{v}, and Ne~\textsc{viii} column densities to predictions
  from equilibrium and nonequilibrium collisional ionization models.
  Observed columns are plotted with discrete symbols at the
  temperature that provides the best fit to the models, which are
  shown with smooth curves. Species corresponding to the symbols and
  curves are indicated in the legend at upper left.  The left column
  shows fits for the component at $v = -317$ km s$^{-1}$ for the
  following models \cite{gnat07}: (a) nonequilibrium ionization in
  plasma that is rapidly and isochorically cooling (i.e., at constant
  density) from an initial temperature of $5 \times 10^{6}$ K, (b)
  same but for isobaric (constant pressure) nonequilibrium cooling
  gas, and (c) collisional ionization equilibrium.  Panels (d) -- (f)
  show the same models for the component at $v = -247$ km s$^{-1}$.
  In both components, Mg~\textsc{x} is not detected, and a 3$\sigma$
  upper limit is shown.}
\end{figure}

We see from Fig.~S3 that the observed N~\textsc{v} and Ne~{\sc viii}
must arise from a more highly ionized and hotter phase than the low
ions.  The strong correlation between the velocity centroids and
profile shapes of the low- and high-ions suggest a physical
connection; the highly ionized species may arise in the
evaporating/ablating surface of the outflowing cool clouds or,
conversely, the cool clouds may be rapidly cooling and condensing out
of the hot outflow.  Some of the intermediate ions (N~\textsc{iv},
O~\textsc{iv}, S~\textsc{v}) may exist in both the photoionized and
hot phases.  To estimate the mass in the highly ionized phase in light
of these possibilities, we use the Ne~{\sc viii}/{\sc N~v} ratio to
constrain the hot-phase gas temperature using three collisional
ionization models \cite{gnat07}, including rapidly cooling models, as
shown in Figure~S5. Then we assume the Ne~{\sc viii} phase has the
same metallicity as the cooler (photoionized) cloud at the same
velocity and adjust $N({\rm H})_{\rm total}$ to match the observed
$N$(Ne~{\sc viii}) and $N$({\sc N~v}).  Figure~S5 shows examples of
the models for the $-317$ and $-247$ km s$^{-1}$ components, including
the isochoric and isobaric nonequilibrium models and the collisional
ionization equilibrium (CIE) model from \cite{gnat07}. Table~S2
summarizes the masses (from the thin-shell model above) and gas
temperatures derived for the hot phase of three of the components; the
lowest masses and temperatures in Table~S2 result from the isochoric
nonequilibrium (rapidly cooling) model, and the highest
masses/temperatures result from assuming CIE.  We find that all three
models provide acceptable fits and imply comparable masses.
Interestingly, when these models are fit based on Ne~{\sc viii}/{\sc
  N~v}, they naturally predict about the right amount of S~\textsc{v}.
The models are also consistent with the upper limits on Mg~\textsc{x}
and the looser constraints on species that are more uncertain due to
moderate saturation (e.g., N~\textsc{iv}, O~\textsc{iv}). We have also
investigated whether the Ne~\textsc{viii} phase could arise in gas
photoionized by AGN flux from the affiliated galaxy.  This model does
not work well -- we cannot match the Ne~\textsc{viii}/N~\textsc{v}
without significantly violating other constraints (other column
densities) from the data.

In principle, the hot gas could have a higher metallicity than the
cool gas.  The hot extraplanar gas of NGC1569, for example, has a
higher metallicity than its underlying {\sc H~i} disk\cite{martin02},
which suggests that chemically enriched hot gas is preferentially
escaping into the halo, as predicted in some galactic wind models
\cite{maclow99}.  However, the high degree of similarity of the low-
and high-ion profiles in the PG1206+459 absorber (Figure~3) argues
against this situation.  If the cool and hot gas had separated
sufficiently so that the hot phase had significantly higher
metallicity, then the cool and hot gas would have different
kinematical properties.  Instead, the low- and high-ion profiles are
remarkably similar. We note that in the central galaxies of
cooling-flow clusters, the metallicity of the hot phase has been
estimated to range from 1$\times$ to 3$\times$ the metallicity of the
cooler phase based single-temperature models fit to X-ray emission
\cite{mcdonald10}, so our assumption of similar metallicites in the
cool and hot phases would be reasonable in this context.  However, as
emphasized in \cite{mcdonald10}, there are substantial systematic
uncertainties in those metallicity estimates, and in some cases the
signal-to-noise is not adequate to support definitive conclusions.
Moreover, the ``cool'' phase metallicity is also based on X-ray
emission data with low spectral resolution, so we do not know if the
cool and hot phases in that study are as well-aligned kinematically as
the UV data presented in this paper.

{\bf Acknowledgements} Based on observations made with the NASA/ESA
Hubble Space Telescope, the MMT, a joint facility operated by the
Smithsonian Astrophysical Observatory and the University of Arizona,
and the LBT, an international collaboration among institutions in the
United States, Italy, and Germany. Support for HST Program number
11741 was provided by NASA through a grant from the Space Telescope
Science Institute, which is operated by the Association of
Universities for Research in Astronomy, Incorporated, under NASA
contract NAS5-26555.  Additional support was provided by NASA grant
NNX08AJ44G. The DEEP2 survey was supported was by NSF grants AST
95-29098, 00-711098, 05-07483, 08-08133, 00-71048, 05-07428, and
08-07630. Funding for SDSS has been provided by the Alfred P. Sloan
Foundation, the Participating Institutions, the National Aeronautics
and Space Administration,the National Science Foundation, the
U.S. Department of Energy Office of Science, the Japanese
Monbukagakusho, and the Max Planck Society.  We thank Chris Churchill
for providing the archival Keck data and the referees for review
comments that significantly improved this paper.  We are also grateful
to the Hawaiian people for graciously allowing us to conduct
observations from Mauna Kea, a revered place in the culture of Hawaii.
The \textit{Hubble Space Telescope} data in this paper is available
from the MAST archive at http://archive.stsci.edu.


\begin{thebibliography}{}
\bibitem{refcolor} The galaxy appears to be red in
  Fig.~\ref{galaxy_montage} due to its redshift; in the rest frame of
  the galaxy, it has a very blue U-B color.
\bibitem{cosmology} For distance calculations, we assume a cold dark
  matter cosmology with $H_{0} = 70$ km s$^{-1}$ Mpc$^{-1}$,
  $\Omega_{\rm m} = 0.30, \Omega_{\Lambda} = 0.70$.
\bibitem{neargalnote} As discussed in the supporting material, the
  yellow galaxy northwest of the QSO (Figure~\ref{galaxy_montage})
  does not have a spectroscopic redshift but is likely to have $z \ll
  0.927$.
\end{thebibliography}
\end{document}